\documentclass[a4paper,fleqn,usenatbib]{mnras}

\usepackage{times}
\usepackage{mathptmx}

\usepackage{graphicx}	
\usepackage{amsmath}	
\usepackage{amssymb}	
\usepackage{xspace}

\usepackage{amsmath} 
\newcommand{\angstrom}{\textup{\AA}}

\newcommand{\xmm}{\textit{XMM-Newton}\xspace}
\newcommand{\mobh}{M_{\rm BH}}
\newcommand{\ledd}{\lambda_{\rm Edd}}
\newcommand{\lxray}{L_{\rm 2-10\,keV}}
\newcommand{\xspec}{{\sc xspec}\xspace}
\newcommand{\FeK}{Fe\,K$\alpha$\xspace}

\title[Broad Fe K$\alpha$ of AGN]{Dependence of the broad Fe K$\alpha$ line on the
physical parameters of AGN}

\author[Liu et al.,]{%
Zhu Liu,$^{1,2}$\thanks{liuzhu@nao.cas.cn(ZL)}
Weimin Yuan,$^{1,2}$
Youjun Lu,$^{3,2}$
Francisco J. Carrera,$^{4}$
Serena Falocco,$^{5}$
\newauthor%
Xiao-Bo Dong$^{6}$
\\
$^1$Key Laboratory of Space Astronomy and Technology, National Astronomical Observatories, Chinese Academy of Sciences, Beijing 100012, China\\
$^2$University of Chinese Academy of Sciences, School of Astronomy and Space Science, Beijing 100049, China\\
$^3$National Astronomical Observatories, Chinese Academy of Sciences, Beijing 100012, China\\
$^4$Instituto de F{\'i}sica de Cantabria (CSIC-UC), Avenida de los Castros, 39005 Santander, Spain\\
$^5$KTH Royal Institute of Technology, Department of Physics  and the Oskar Klein Centre, AlbaNova, SE-106 91 Stockholm, Sweden\\
$^6$Yunnan Observatories, Chinese Academy of Sciences, Kunming, Yunnan 650011, China; Key Laboratory for the Structure and Evolution\\$^\text{ }$of Celestial Objects, Chinese Academy of Sciences, Kunming, Yunnan 650011, China\\
}

\date{Accepted XXX. Received YYY; in original form ZZZ}

\pubyear{2016}

\begin{document}
\label{firstpage}
\pagerange{\pageref{firstpage}--\pageref{lastpage}}
\maketitle

\vspace{10mm}
\begin{abstract}
In this paper, the dependence of the broad \FeK line on the physical parameters of AGN, such as the black hole mass $\mobh$, accretion rate (equivalently represented by Eddington ratio $\ledd$), and optical classification, is investigated by applying the X-ray spectra stacking method to a large sample of AGN which have well measured optical parameters. A broad line feature is detected ($>3\sigma$) in the stacked spectra of the high $\ledd$ sub-sample ($\log\ledd>-0.9$). The profile of the broad line can be well fitted with relativistic broad line model, with the line energy consistent with highly ionized Fe K$\alpha$ line (i.e. \ion{Fe}{xxvi}). A model consisting of multiple narrow lines cannot be ruled out, however. We found hints that the Fe K line becomes broader as the $\ledd$ increases. No broad line feature is shown in the sub-sample of broad-line Seyfert 1 (BLS1) galaxies and in the full sample, while a broad line might be present, though at low significance, in the sub-sample of narrow-line Seyfert 1 (NLS1) galaxies. We find no strong dependence of the broad line on black hole masses. Our results indicate that the detection/properties of the broad \FeK line may strongly depend on $\ledd$, which can be explained if the ionization state and/or truncation radius of the accretion disc changes with $\ledd$. The non-detection of the broad line in the BLS1 sub-sample can be explained if the the average EW of the relativistic Fe K$\alpha$ line is weak or/and the fraction of sources with relativistic Fe K$\alpha$ line is small in BLS1 galaxies.
\end{abstract}





\section{Introduction}\label{sec:intro}

Observational evidences of a broad \FeK line feature are found in the X-ray spectra of some active galactic nuclei (AGN), such as MCG-6-30-15 \citep{tanaka:1995, fabian:2002, miniutti:2007}, NGC 3516 \citep{turner:2002, markowitz:2006}, 1H 0707-495 \citep{fabian:2009} and others \citep{miller:2007, nandra:2007}. The broad \FeK line is generally believed to originate from the inner region of the accretion disc via the K-shell fluorescence process, and to be broadened due to the Doppler boosting, gravitational redshift and the transverse Doppler effect \citep{fabian:1989}. The energy at which the ``red'' wing of the broad line truncates is directly linked to the inner radius of the accretion disc that is commonly thought to be at the innermost stable circular orbit (ISCO). The spin of the black hole, which is related to the ISCO through a monotonic relation \citep{bardeen:1972}, can be inferred by modeling the broad \FeK line profile \citep{brenneman:2006, dauser:2010}.

Besides its mass, an astrophysical black hole is completely characterized by its spin. The distribution of the black hole spin may yield important insights into the growth history and accretion process of supermassive black holes (SMBH). For instance, models in which SMBH growth is dominated by BH-BH mergers can lead to a bimodal distribution with one peaked at 0 and the other located at $\sim0.7$, whereas growth via gas accretion predicts a rapidly spinning or a slowly spinning population depending upon whether the BHs gain their masses via prolonged accretion or chaotic accretion \citep{moderski:1996, volonteri:2005}. Moreover, spin also determines the radiative efficiency, which is the mass-to-energy conversion efficiency, and thus influences how efficiently BHs accrete mass during the accretion phase. Black hole spin can also be a potent energy source, and may drive the powerful relativistic jets that are seen from many BH systems through the Blandford-Znajek mechanism \citep{blandford:1977}.

There have been several studies for charactering the broad \FeK line and measuring the spin of SMBH based on relativistic reflection spectra in the literature \citep[e.g.][]{nandra:2007, de-la-calle-perez:2010, patrick:2012, walton:2013}. However, a significantly broad \FeK line is detected in only $<50$\, per cent of the sources in previous studies with different samples. In total, detections of the broad \FeK line are reported in the X-ray spectra of $\sim46$ AGNs. Among those, about 22 sources have reliable spin measurements \citep{reynolds:2013, brenneman:2013, reynolds:2014}, making it difficult to draw any robust statistical inferences on the distribution of BH spin. 

The reason for the lack of apparent relativistic broad \FeK line in the X-ray spectra of some AGN is still unclear. Observationally, one possible explanation is the low signal-to-noise (S/N) of the X-ray data for the majority of AGN. As demonstrated in \citet{mantovani:2014}, the broad \FeK line is revealed in the composite X-ray spectrum of IC 4329A observed by {\it Suzaku}, while it is absent in each individual observation due to the low S/N of the data. Indeed, to determine accurately the continuum spectrum and to reveal unambiguously the \FeK line profile, a large number of X-ray photons collected at high energies is required \citep{de-la-calle-perez:2010, nandra:2007, mantovani:2014}. Theoretically, the strength of the line is a function of the geometry of the accretion disc, which determines the solid angle subtended by the reflecting matter as seen by the X-ray source. It also depends on the elemental abundances of the reflecting matter, the inclination angle at which the reflecting surface is viewed, and the ionization state of the surface layers of the disc. \citet{bhayani:2011} analyzed a sample of 11 Seyfert galaxies observed by XMM-Newton that appear to be missing a broad \FeK line. They argued that the lack of apparent relativistic \FeK line can be explained if this feature becomes indistinguishable from the underlying continuum, as a result of a combination of several effects, e.g. blending and Comptonization in an ionized disc, strong relativistic effects and, in some cases, a high disc inclination. Calculations have also shown that the ionization state of the accretion disc, which will affect the observed line energy as well as the line equivalent width (EW) of the broad line \citep{matt:1993, ross:1993, nayakshin:2000}, depends strongly on the accretion rate. Thus a correlation between the properties of the broad \FeK line and the accretion rate is expected. Such a correlation has been reported in \citet{inoue:2007}, although a relativistic line is statistically not required by their data.

In order to investigate the correlation between the properties of the broad \FeK line and the physical parameters of AGN, a homogeneously selected sample of AGN with well measured optical parameters is preferred. The virial BH mass $\mobh$ of AGN can be estimated through the empirical relations using the emission line widths and continuum luminosities \citep{kaspi:2005, shen:2013}. The Eddington ratio $\ledd$ is given by ${\ledd} = L_{\rm bol}/L_{\rm Edd}$, where the $L_{\rm Edd}$ is defined as $L_{\rm Edd} = 1.26\times10^{38}(\mobh/M_{\odot})$. The bolometric luminosity $L_{\rm bol}$ is usually estimated from the integration of the broadband spectral energy distribution (SED) or from a single band assuming a bolometric correction. The $\mobh$ and $\ledd$ for individual sources or small samples in previous studies (e.g. \citealt{inoue:2007}) were derived using different data analysis methods, BH mass estimation formalisms and/or heterogeneous data sets. Because of the lack of large homogeneously selected samples with high S/N X-ray data and well measured optical parameters, it is still unclear how is the appearance/property of the broad \FeK line dependent on the physical properties of AGN.

X-ray spectral stacking is an effective way to obtain composite spectra with very high S/N. A broad relativistic Fe K$\alpha$ line is found in the stacked spectra of the Lockman Hole field using \xmm observations \citep{streblyanska:2005}. However, while a narrow line is significantly detected, the broad line is not clearly seen in the stack spectra of different samples in previous studies \citep{corral:2008, chaudhary:2010, chaudhary:2012, iwasawa:2012a, falocco:2013}. The relativistic Fe K$\alpha$ line is detect at $6\sigma$ using a sample with high S/N observed with \xmm in \citet{falocco:2014}. They pointed out that the low average S/N of the spectra, which make the continuum not to be well determined, can explain the low significance of the broad line in previous works. The non-detection of the broad line in the average spectra, which are obtained using samples including sources with diverse properties, can also be interpreted if the property of the relativistic line is highly dependent on one or more physical parameters of AGN. By applying the same rest-frame X-ray spectral stacking method \citep{corral:2008, falocco:2012} to a sample of NLS1 galaxies, \citet{liu:2015} found that there exists a prominent broad \FeK line in the composite X-ray spectrum. They suggested that  broad \FeK line is perhaps common in AGN with high $\ledd$ (e.g. NLS1 galaxies). The average properties of the relativistic Fe K$\alpha$ line for a uniform selected BLS1 galaxies sample have not been explored yet, even though a broad \FeK line is detected in several BLS1 galaxies.

In this paper, using a large homogeneously selected sample of AGN which have well measured $\mobh$ and $\ledd$, we investigate how the detection/property of the broad \FeK line depends on the physical parameters, such as the $\mobh$, $\ledd$, and optical classification of AGN, by means of X-ray spectral stacking. We use the cosmological parameters $H_\mathrm{0}=70\,\mathrm{km\,s^{-1}\,Mpc^{-1}}$, $\Omega_\mathrm{M}=0.27$, and $\Omega_\mathrm{\Lambda}=0.73$. All quoted errors correspond to the 90 per cent confidence level for one interesting parameter, unless specified otherwise.


\section{Sample and data reduction}\label{sec:sample}

\begin{figure}
  \includegraphics{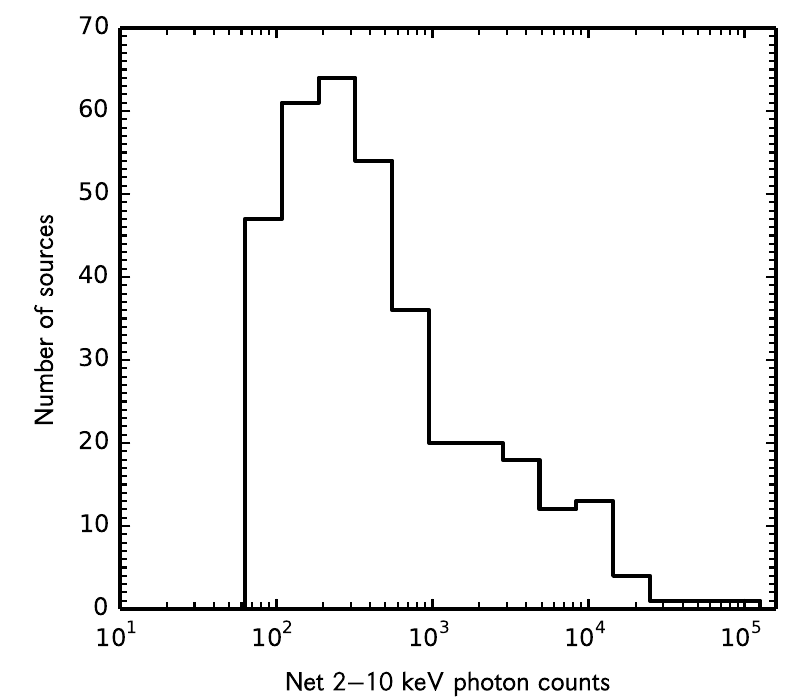}
  \caption{\label{fig:ph_dist}Distribution of the rest-frame $2-10\,\mathrm{keV}$ net photon counts.}
\end{figure}

We compiled a sample of type-1 AGN that have \xmm observations from an optical selected broad line AGN catalogue \citep{dong:2012}. This optical catalogue, being the parent sample of the intermediate mass black hole (IMBH) AGN sample of \citet{dong:2012}, consists of 8862 objects at $z\leq 0.35$ that were homogeneously selected from the Sloan Digital Sky Survey Data Release 4 (SDSS-DR4; \citealt{adelman-mccarthy:2006}). The sources in the optical catalogue were matched to the third generation \xmm Serendipitous Source Catalogue (3XMM-DR5) using a searching radius of 5\,arcsec. We selected those having more than 60 net source counts detected with either the EPIC pn or the combined MOS detectors in the rest-frame 2-10\,keV band. In total, the sample includes 156 sources that have 212 \xmm observations. Some sources have multiple observations; for them we do not find significant variability, and each of their observations is treated as an independent source. The distribution of rest-frame $2-10\,\mathrm{keV}$ photon counts is shown in Fig. \ref{fig:ph_dist}.

\subsection{The optical data analysis}\label{subsec:optical}

\begin{figure*}
\begin{center}
\includegraphics[width=1.0\columnwidth]{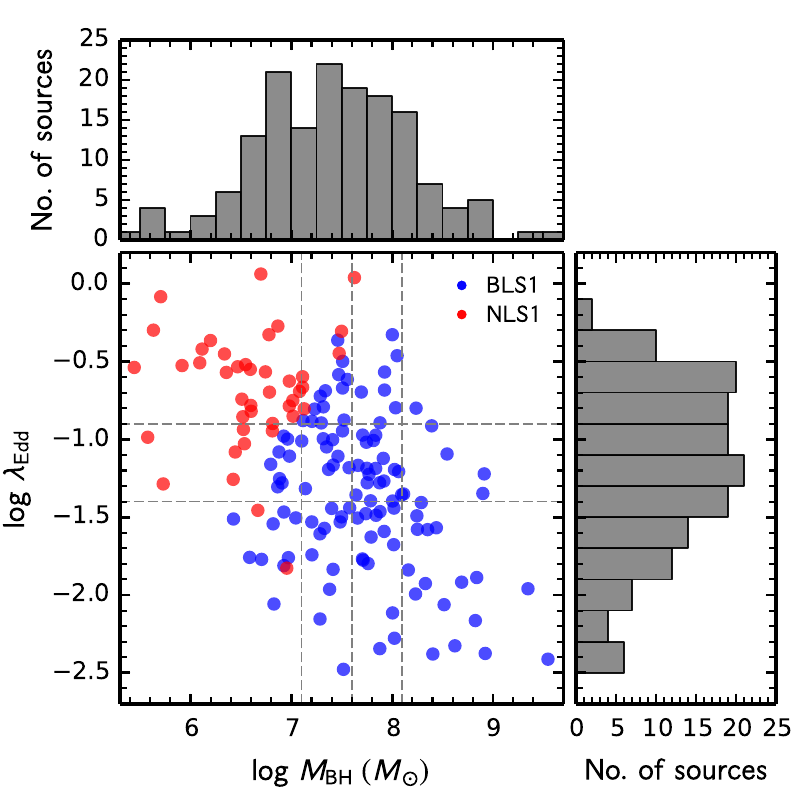}
\includegraphics[width=1.0\columnwidth]{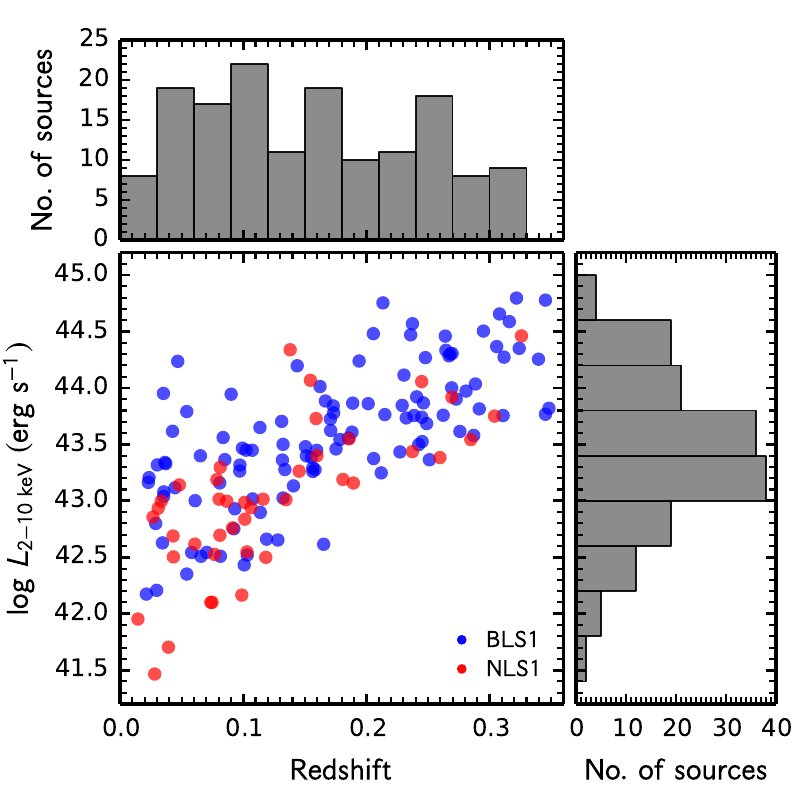}
\caption{Left: The main panel shows the distribution of the sources in the $\mobh-\ledd$ plane. The top and right panels show the distributions of the $\mobh$ and $\ledd$ for the sources in the sample, respectively. The dashed lines mark the boundaries of sub-samples defined in Section\ref{subsec:sub-samples}. Right: the same as the left plot, but for the $L_\mathrm{2-10\,keV}$ versus redshift.}\label{fig:mbh_edd}
\end{center}
\end{figure*}

The details of the optical data analysis can be found in \citet{dong:2012}. The main procedures to fit the continua and emission line profiles are summarized here. Many SDSS spectra have a significant contribution from host galaxy starlight, because of a relatively large fiber aperture of 3\,arcsec in diameter. To remove the starlight, six synthesized galaxy spectral templates \citep{lu:2006} are used to model the host galaxy starlight \citep{zhou:2006}. Two separate sets of analytic templates \citep{veron-cetty:2004, dong:2008} are adopted to model the narrow-line and broad-line \ion{Fe}{ii} emission, respectively. The profiles of all the other emission lines are fitted using the method described in detail in \citet{dong:2005}. Basically, each line (narrow or broad) is fitted incrementally with as many Gaussians as statistically justified. The formalism presented by \citet{greene:2007}, which makes use of the luminosity and FWHM of the broad H$\alpha$ line, is adopted to calculate the $\mobh$ for the sources. The bolometric luminosity is estimated by assuming $L_\mathrm{bol}=9.8\lambda L_\lambda(5100\angstrom)$ \citep{mclure:2004}, while $\lambda L_\lambda(5100\angstrom)$ is calculated from the H$\alpha$ luminosity \citep{greene:2005}. The distributions of the $\mobh$ and $\ledd$ for the 156 sources in the sample of the present work are shown in the left of Fig. \ref{fig:mbh_edd} (top panel: $\mobh$; right panel: $\ledd$). In the right of Fig. \ref{fig:mbh_edd} we show the distribution of the sources in the $L_\mathrm{2-10\,keV}-\mathrm{redshift}$ plane.

\subsection{\label{subsec:sub-samples} Sub-samples}

We define our sub-samples on the basis of their physical properties, i.e. a) four $\mobh$ sub-samples: MBH1 sample ($\log\mobh<7.1$); MBH2 ($7.1<\log\mobh<7.6$); MBH3 ($7.6<\log\mobh<8.1$) and MBH4 sample ($\log\mobh>8.1$), b) three $\ledd$ sub-samples: low $\ledd$ sample ($\log\ledd<-1.4$); medium $\ledd$ ($-1.4<\log\ledd<-0.9$) and high $\ledd$ ($\log\ledd>-0.9$), and c) two sub-samples with different optical classifications: the NLS1 (defined as $\mathrm{FWHM_{H\alpha}}<2000\,\mathrm{km\,s^{-1}}$, $\mobh<2.0\times10^6\,M_{\sun}$) and BLS1 ($\mathrm{FWHM_{H\alpha}}>2000\,\mathrm{km\,s^{-1}}$) sub-samples. Sources with $\mobh<2.0\times10^6\,M_{\sun}$ (e.g. intermediate mass black holes, IMBH, see \citealt{greene:2007, dong:2008}), which may have different properties from NLS1 \citep{ai:2010a}, are excluded from our NLS1 sub-sample. Except for the NLS1 and BLS1 sub-samples, all the other sub-samples are selected in a way that the stacked spectra have similar S/N. The details of each sub-sample can be found in Table \ref{tab:sub_samples}.

\begin{table}
\bgroup
\setlength\tabcolsep{2.8pt}
\begin{center}
\caption{\label{tab:sub_samples}Properties of each sub-sample}
\begin{tabular}{lcccccccr}\hline
Sample  & $\tilde{z}$ & $\tilde{L}_{\rm{2-10~keV}}$ & $\tilde{M}_{\rm{BH}}$ &
$\tilde{\lambda}_{\rm{Edd}}$ & \#src & \#spec & $\tilde{\mathrm{cts}}$ & cts \\
                & (1)  & (2)   & (3)  & (4)   & (5) &  (6)  &  (7) & (8) \\\hline
MBH1            & 0.08 & 42.84 & 6.60 & -0.92 & 78 & 117 & 283 & 125292 \\
MBH2            & 0.15 & 43.47 & 7.38 & -0.95 & 49 &  65 & 183 & 185047 \\
MBH3            & 0.18 & 43.74 & 7.84 & -1.27 & 62 &  95 & 389 & 181356 \\
MBH4            & 0.19 & 44.17 & 8.37 & -1.58 & 54 &  76 & 713 & 188965 \\\hline
Low $\ledd$     & 0.13 & 43.39 & 7.84 & -1.71 & 78 & 113 & 251 & 190840 \\
Medium $\ledd$  & 0.14 & 43.59 & 7.78 & -1.09 & 88 & 139 & 439 & 219270 \\
High $\ledd$    & 0.16 & 43.49 & 7.08 & -0.60 & 77 & 101 & 360 & 270550 \\\hline
NLS1            & 0.10 & 43.01 & 6.68 & -0.69 & 50 &  60 & 247 & 57688\\
BLS1            & 0.16 & 43.69 & 7.81 & -1.31 &179 & 266 & 382 & 613304\\\hline
Full sample     & 0.14 & 43.48 & 7.52 & -1.09 &243 & 353 & 329 & 680660\\\hline
\end{tabular}
\parbox[]{83mm}{The columns are: (1) median redshift; (2) median 2-10~keV luminosity in unit erg~s$^{-1}$; (3) median $\mobh$; (4) median $\ledd$; (5) number of sources, some sources have multiple observations, each of their observation is treated as an independent source; (6) number of spectra, pn and MOS spectra are counted independently; (7) median rest-frame $2-10\,\mathrm{keV}$ photon counts; (8) total rest-frame $2-10\,\mathrm{keV}$ net photon counts}
\end{center}
\egroup
\end{table}


\subsection{The X-ray data reduction}\label{subsec:xray}

The Observation Data Files (ODFs) of all the \xmm observations are obtained from the \xmm public archive data. The ODFs are reduced using the \xmm Sciences Analysis System (\textsc{sas}) version 14.0 \citep{gabriel:2004}. We use the \textsc{sas} tasks \textsc{emchain} and \textsc{epchain} to produce the event lists for the European Photon Imaging Camera (EPIC) pn \citep{struder:2001} and MOS \citep{turner:2001} detectors, respectively. Flaring background periods are identified and excluded using the \textsc{sas} task \textsc{espfilt}. The source spectra are extracted from each individual observation using a circular region, of which the radius is in the range of 15-40\,arcsec according to the S/N and the off-axis angle of the detection. X-ray events with pattern $\leq12$ for MOS and $\leq4$ for pn are used to extract the X-ray spectra. The background spectra of the EPIC MOS camera are extracted from a source-free concentric annulus or circles with roughly the same off-axis angle located on the same CCD chip as the sources, while the background spectra of the pn camera are extracted from circular regions centered at the same CCD read-out column as the source positions. The {\sc sas} tasks {\sc rmfgen} and {\sc arfgen} are used to generate the redistribution matrix and the ancillary file. In order to increase the S/N, the two MOS spectra of each observation are combined when are available. 


\section{X-ray spectral stacking of the sample}

Spectral stacking is an effective way to obtain a composite spectrum with very high signal-to-noise (S/N) for a certain sample selected on the basis of their physical properties. It is useful to reveal spectral features in the composite spectrum, such as the broad Fe\,K$\alpha$ line, which can be too weak to be detected in individual sources. 

\subsection{Spectral stacking method}

In general, the measured X-ray spectra of different sources and instruments, which are a result of the convolution of the source spectra with the instrumental responses, cannot be co-added directly. In this work, we adopted a rest-frame stacking method presented in \citet[][see also \citealt{falocco:2012}]{corral:2008} to stack the X-ray spectra of the sources in our sample. The main procedures are outlined here.
\begin{enumerate}
    \item Determining the source continuum: For each source, the observed pn and MOS (if both are available) are jointly fitted with a simple power-law modified by Galactic and possible intrinsic absorption. The spectrum below 2$~$keV is excluded to avoid possible contamination from the soft X-ray excess. The 5-7$~$keV energy range is ignored to exclude contribution from potential \FeK emission line features. The Galactic absorption column density is fixed at the Galactic value given by \citet{kalberla:2005} for each source. In this way, the best fitted continuum model is obtained and the $\lxray$ for each source is calculated.
    \item Unfolding the spectrum: The source spectrum before entering the telescope can be reconstructed with the instrumental effects eliminated by unfolding the observed spectrum with the calculated conversion factors from the best-fit model. This is done by using the \xspec command  \textsc{eufspec}.
    \item Rescaling: The unfolded spectra are corrected for both the Galactic and intrinsic absorption effects and de-redshifted to the source rest frame. Following \citet{corral:2008}, we rescale the spectra so that they have the same rest frame 2-5$~$keV fluxes.
    \item Rebin: The de-redshifted and rescaled rest-frame spectra should be rebinned before stacking. To construct a new, unified bin scale, we group the rescaled spectrum of each source with a bin width of 100\,eV.
    \item Stacking: We obtain the stacked spectrum by averaging the rebinned and rescaled spectra using the simple arithmetic mean. The errors are calculated using the propagation of errors.
\end{enumerate}

\subsection{The emission line feature and its significance}\label{subsec:simulations}

\begin{figure*}
    \begin{center}
    \includegraphics[width=\linewidth]{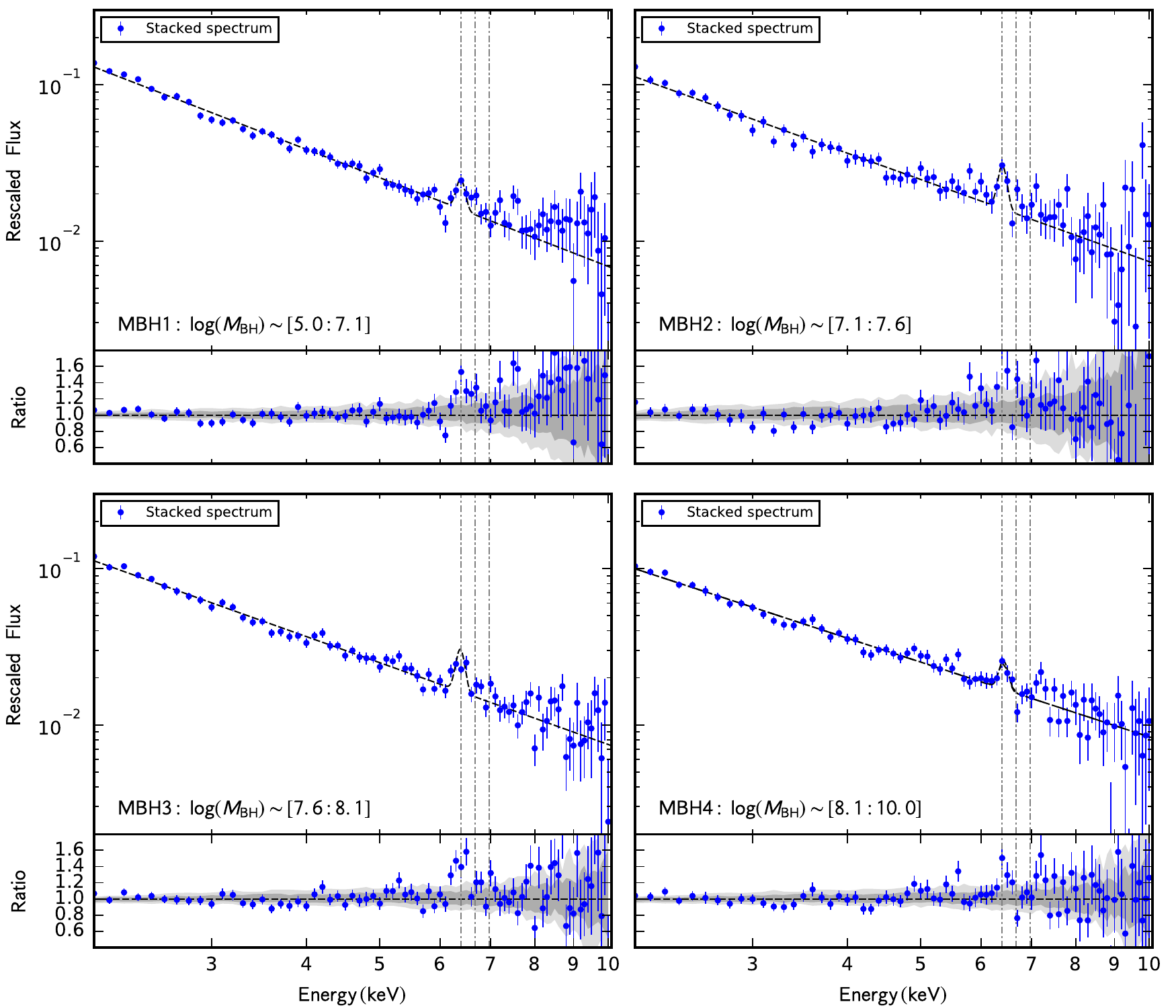}
    \caption{\label{fig:mbh_subs}The stacked spectra of the $\mobh$ sub-samples, with the best-fitting power-law continua (black dashed line) are shown in the {\it upper} plot of each panel. The profile of an unresolved line at 6.4\,keV superposing the power-law is also shown for illustration purposes. The ratios of the stacked spectra to the best-fitting power-law continua are shown in the {\it lower} plots of each panels. The $1\sigma$ and $2\sigma$ confidence intervals are marked as dark and light shaded areas, respectively. The gray dashed vertical lines mark the energies of the 6.4\,keV, 6.67\,keV and 6.97\,keV emission lines.}
\end{center}
\end{figure*}

\begin{figure}
    \begin{center}
    \includegraphics[width=1.0\columnwidth]{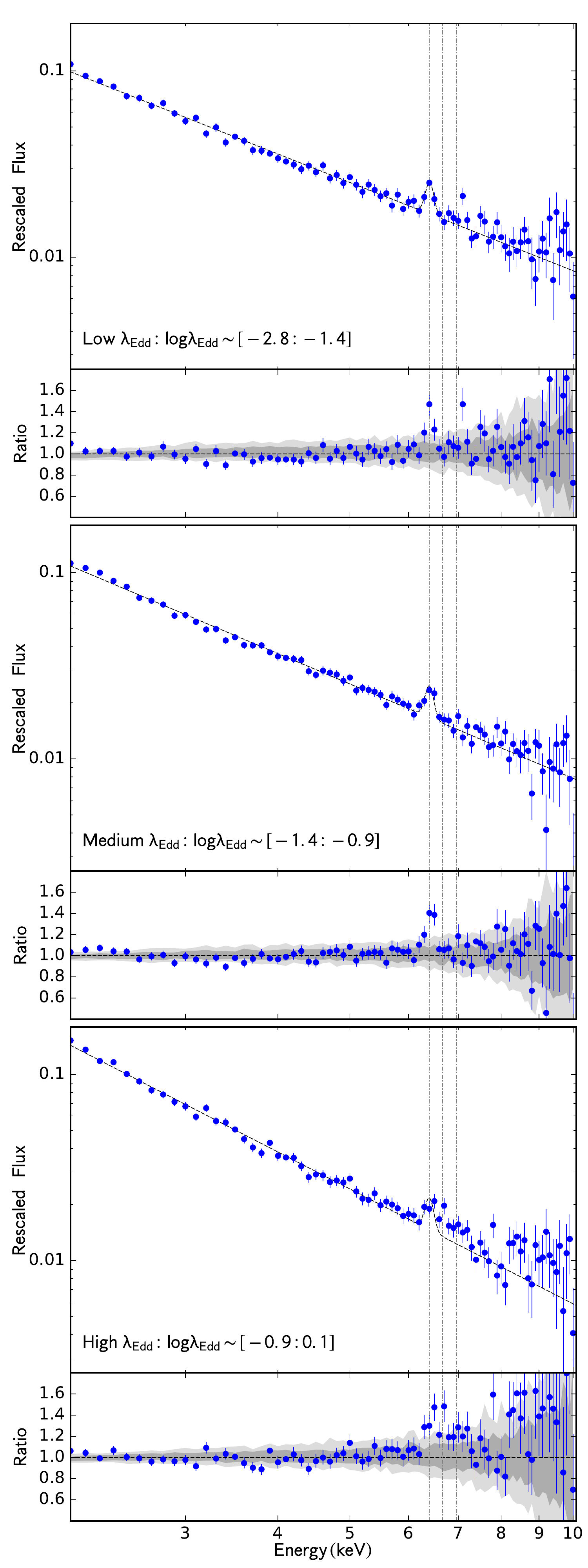}
    \caption{\label{fig:edd_subs}Same as Fig. \ref{fig:mbh_subs}, but for the $\ledd$ sub-samples. A broad line feature can be seen in the $6-7\,\mathrm{keV}$ energy range of the high $\ledd$ sub-sample.}
\end{center}
\end{figure}

\begin{figure*}
    \begin{center}
    \includegraphics[width=\linewidth]{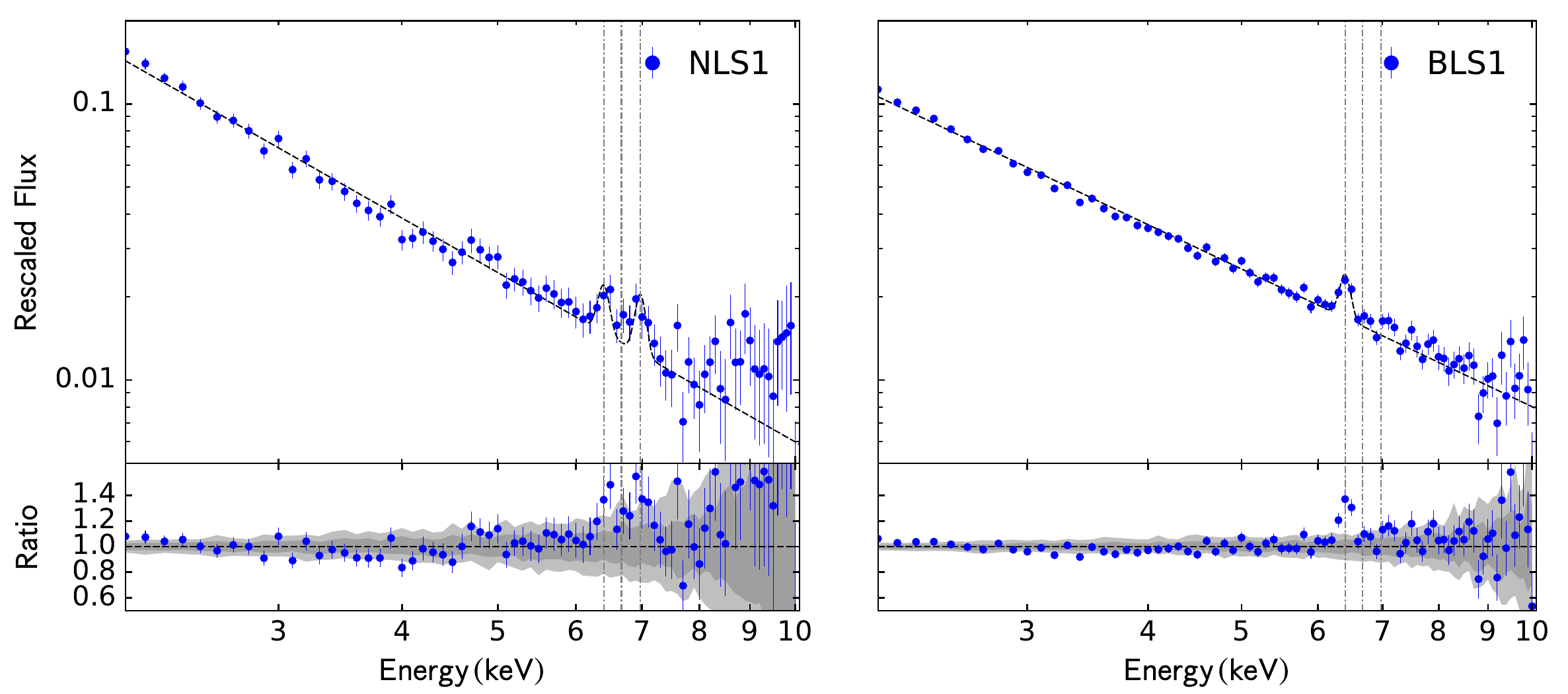}
    \caption{\label{fig:nl_bl}Same as Fig. \ref{fig:mbh_subs}, but for the nls1-bls1 sub-samples. The profiles of an unresolved line at 6.4\,keV and 6.97\,keV is also shown in the NLS1 sub-sample.}
\end{center}
\end{figure*}

The rest-frame 2-10\,keV stacked spectra for each sub-subsample are shown in Figs.\,\ref{fig:mbh_subs} (the $\mobh$ sub-samples), \ref{fig:edd_subs} (the $\ledd$ sub-samples) and \ref{fig:nl_bl} (the NLS1-BLS1 sub-samples).  The stacked spectra are marked with blue points. The ratio of the data to a best-fitting power-law continuum is also included in each plot (bottom panel). An emission line feature peaking at around 6.4\,keV, which must be the neutral \FeK line, is clearly shown in all the stacked spectra. Moreover, indication of a broad emission line feature in the 6-7\,keV energy range is also found in the stacked spectra of the high $\ledd$. No significant absorption feature is found in the stacked spectra.

We carry out simulations to construct the confidence intervals for the underlying continua of the composite spectra, as well as to estimate the significances of the \FeK emission feature. Following \citet[][see also \citealt{corral:2008, falocco:2012}]{liu:2015}, we simulate 100 spectra using \xspec for each of the sources in our sample, using the best-fitting continuum spectrum model. The exposure time, auxiliary and response matrix files for the sources in the sample are used in the simulations. By applying the same stacking method to the simulated spectra, we generate 100 stacked spectra for each sub-sample. We roughly estimate the $1\sigma$ and $2\sigma$ confidence intervals by calculating the 2ed, 16th, 84th, 98th percentile values in each energy bin, which are shown as gray and dark gray shadowed areas in Fig.\,\ref{fig:mbh_subs}-\ref{fig:nl_bl}. It is clear that the 6.4\,keV emission features are detected at 2\,$\sigma$ level in all the stacked spectra of the sub-samples, while indications of the broad emission features are shown in the stacked spectra of the high $\ledd$ and NLS1 sub-samples. As shown in Fig. \ref{fig:edd_subs}, at least 8 adjacent data points in the $6-7\,\mathrm{keV}$ region of the high $\ledd$ sub-sample fall at or out of the $2\sigma$ confidence levels. The probability that the broad line profile arises from statistical fluctuations is small, i.e. $<4\times10^{-11}$ ($P<0.05^8$).

We further investigate whether the profile of a narrow line can be significantly affected by the instrument effects and the stacking method. To examine this, we generate a series of observed spectra with high S/N using a model consisting of a power-law and an unresolved Gaussian line ($\sigma=1\,\mathrm{eV}$, $\mathrm{EW}=200\,\mathrm{eV}$). The central energy of the line is fixed at the Fe K emission line region, e.g. $6.4-6.97\,\mathrm{keV}$ ($\mathrm{\ion{Fe}{I}}-\mathrm{\ion{Fe}{XXVI}}$). We also simulate spectra using the same method but with the line energy fixed at values between 3 and 9\,keV, in steps of 1\,keV. The same stacking method is then applied to the simulated spectra. We find that the line width of the $6.4-6.97\,\mathrm{keV}$ Gaussian profile in the simulated stacked spectrum will be $\sigma\approx90\,\mathrm{eV}$. These results are consistent with the conclusions reached by previous studies \citep{corral:2008, iwasawa:2012a, falocco:2013, falcke:2004, liu:2015}. To account for the line broadening effect, we convolve the models in the spectral fits with a Gaussian smoothing (\texttt{gsmooth} in \xspec, see Section \ref{subsec:modeling}). For each sub-sample the line widths at energies between 3 and 9\,keV are estimated from the simulated average spectra. Following \citet{falocco:2014}, we compute the line width as a power-law function of the line energy. The two parameters in the \texttt{gsmooth}, the line width at 6\,keV and the power-law index, can be found in Table \ref{tab:fit_subs} ($\sum_\mathrm{6\,keV}$ and $\alpha$) for each sub-sample. In addition to the Fe K emission lines, unidentified line features are also shown in some of the stacked spectra. In most cases they are present in the $E>7\,\mathrm{keV}$ energy band where the S/N is low. These line features can be due to systematical uncertainties of the data.

\subsection{X-ray spectra modeling}
\label{subsec:modeling}

We use {\sc xspec} (version 12.8) to fit all the stacked spectra. We use the \textsc{ftools} task \textsc{flx2xsp} to convert the flux spectra into fits format which can be fitted using {\sc xspec}. We define a base-line model consisting of a power-law continuum and a Gaussian line with the line energy as free parameter, i.e. \texttt{gsmooth$\otimes$(po+gaussian)} in \xspec. The line width is fixed at 1\,eV if its best-fitting value is consistent with an unresolved line (e.g. close to 0\,eV). The significance of the Gaussian component is assessed by comparing the change in $\chi^2$ between the best-fitting $\chi^2$ value with line flux as free parameter and that with the line flux fixed at 0. The $\Delta\chi^2$ for each sub-sample is list in Table \ref{tab:fit_subs}. The line is significantly detected ($>99.7\%$) in all sub-samples. The slightly lower significance in NLS1 ($\Delta\chi^2=9.9$) may be due to the weak EW of the narrow line (see Section \ref{subsubsec:nlbl_subs}) as well as the much lower S/N (see Table \ref{tab:sub_samples}). 

\begin{table*}
\bgroup
\begin{center}
\caption{\label{tab:fit_subs}Spectral fit with the baseline model for each sub-sample}
\begin{tabular}{@{}lccrrcccccc@{}}
\hline
  \multicolumn{1}{@{}l}{Sub-sample} &
  \multicolumn{1}{c}{$\Gamma$} &
  \multicolumn{1}{c}{$E\,\mathrm{(keV)}$} &
  \multicolumn{1}{c}{$\sigma\,\mathrm{(eV)}$} &
  \multicolumn{1}{c}{$\mathrm{EW\,(eV)}$} &
  \multicolumn{1}{c}{$\sum_\mathrm{6\,keV}$} &
  \multicolumn{1}{c}{$\alpha$} &
  \multicolumn{1}{c}{$\chi^2_\mathrm{s}/\mathrm{d.o.f}$} &
  \multicolumn{1}{c}{$\Delta\chi^2_1$} &
  \multicolumn{1}{c}{$\Delta\chi^2_2$} &
  \multicolumn{1}{c@{}}{$\chi^2_\mathrm{d}/\mathrm{d.o.f}$}\\
  \multicolumn{1}{@{}l}{(1)} &
  \multicolumn{1}{c}{(2)} &
  \multicolumn{1}{c}{(3)} &
  \multicolumn{1}{c}{(4)} &
  \multicolumn{1}{c}{(5)} &
  \multicolumn{1}{c}{(6)} &
  \multicolumn{1}{c}{(7)} &
  \multicolumn{1}{c}{(8)} &
  \multicolumn{1}{c}{(9)} &
  \multicolumn{1}{c}{(10)} &
  \multicolumn{1}{c}{(11)} \\\hline
MBH1           & $1.87\pm0.05$ &  $6.41\pm0.06$ & $1  $               & $116_{-45}^{ +49}$ & 79 & 0.42 & $86.85/76$ & 17.95 &   --- & 81.26/74\\[0.5mm]
MBH2           & $1.74\pm0.07$ &  $6.40\pm0.04$ & $1  $               & $186_{-71}^{ +74}$ & 81 & 0.41 & $90.99/76$ & 19.26 &   --- &   ---   \\[0.5mm]
MBH3           & $1.73\pm0.05$ &  $6.38\pm0.06$ & $106_{ -61}^{ +79}$ & $179_{-67}^{ +68}$ & 80 & 0.40 & $90.35/75$ & 30.88 &  4.7  &   ---   \\[0.5mm]
MBH4           & $1.59\pm0.05$ &  $6.42\pm0.05$ & $1  $               & $109_{-45}^{ +44}$ & 82 & 0.37 & $88.11/76$ & 17.52 &   --- &   ---   \\[0.5mm]\hline
Low $\ledd$    & $1.58\pm0.05$ &  $6.40\pm0.04$ & $1  $               & $ 98_{-33}^{ +32}$ & 80 & 0.39 & $74.71/76$ & 25.64 &   --- &   ---   \\[0.5mm]
Medium $\ledd$ & $1.68\pm0.03$ &  $6.42\pm0.04$ & $58_{ -58}^{ +67}$  & $113_{-34}^{ +35}$ & 80 & 0.41 & $77.78/75$ & 37.39 &  2.9  &   ---   \\[0.5mm]
High $\ledd$   & $2.05\pm0.06$ &  $6.62\pm0.16$ & $>219$              & $325_{-121}^{+128}$& 82 & 0.41 & $81.63/75$ & 33.35 & 14.52 & 77.99/73\\[0.5mm]\hline
NLS1           & $2.0 \pm0.06$ &  $6.46\pm0.08$ & $1  $               & $102_{-52}^{ +52}$ & 81 & 0.42 & $87.48/76$ &  9.9  &   --- & 76.15/74\\[0.5mm]
BLS1           & $1.66\pm0.02$ &  $6.41\pm0.03$ & $57_{ -57}^{ +61}$  & $100_{-25}^{ +27}$ & 80 & 0.39 & $102.8/75$ & 60.63 &  1.0  &   ---   \\\hline
\end{tabular}
\parbox{148mm}{(1) Different sub-samples; (2) The best-fitting photon index; (3) The best-fitting line energy of the Gaussian component; (4) The best-fitting line width of the Gaussian component, it is fixed at 1\,eV if the best-fitting value is consistent with an unresolved line (e.g. close to 0.0); (5) The EW of the Gaussian component; (6) The line dispersion at 6\,keV in the \texttt{gsmooth} model; (7) The index of the power-law function in the \texttt{gsmooth} model; (8) $\chi^2/\mathrm{d.o.f}$ of the baseline model fit; (9) Change in $\chi^2$ if the line flux of the Gaussian component is fixed at 0.0; (10) Change in $\chi^2$ if the line width of the Gaussian component is fixed at 0\,eV; (11) $\chi^2/\mathrm{d.o.f}$ if a second Gaussian component is added.\\
}
\end{center}
\egroup
\end{table*}


\subsubsection{\label{subsubsec:mbh_subs}The $\mobh$ sub-samples}

The stacked spectra of the MBH2, MBH3 and MBH4 sub-samples can be well fitted with the base line model, while a second Gaussian component is added to the baseline model to account for the ionized emission line features peaked at around 6.7\,keV in the stacked spectra of the MBH1 sub-sample. The ionized line in the stacked spectrum of MBH1 is unresolved, thus the line width is fixed at $1\,\mathrm{eV}$. The best-fitting line energy is $6.66^{+0.10}_{-0.11}\,\mathrm{keV}$, corresponding to the \ion{Fe}{xxv} line. The equivalent width of this component is $63^{+47}_{-44}\,\mathrm{eV}$. The line width and the line energy of the Gaussian component in the baseline model are consistent with an unresolved neutral Fe K$\alpha$ line in all but the MBH3 sub-samples (see Table \ref{tab:fit_subs}). The line width of the Gaussian component is $106_{ -61}^{ +79}\,\mathrm{eV}$ in the stacked spectra of MBH3. This may be an indication a broad line profile ($\sim2\sigma$, $\Delta\chi^2=4.7$ if fixed the line width at 0\,eV). The best-fitting photon index of the power-law continuum for MBH4 is quite flat, i.e. $\Gamma=1.59\pm0.05$. As shown in Fig.\,\ref{fig:mbh_edd}, the majority of the sources in the MBH4 sub-samples have low $\ledd$ ($\tilde{\lambda}_{\rm{Edd}}=-1.58$). Thus the flat power-law continuum in MBH4 is expected and is in agreement with the value obtained using the empirical $\Gamma-\ledd$ relation \citep[e.g.][]{risaliti:2009, brightman:2013}.

\subsubsection{\label{subsubsec:edd_subs}The $\ledd$ sub-samples}

The baseline model can fit the stacked spectra well (see Table \ref{tab:fit_subs}). It clearly shows a positive correlation between $\Gamma$ and $\ledd$. This is consistent with the results found in previous studies \citep[e.g.][]{lu:1999, shemmer:2006, risaliti:2009, brightman:2013}, though a flatter relation at high $\ledd$ (e.g. $\geq1.0$) has been suggested by \citet[][see also \citealt{kamizasa:2012}]{ai:2010a}. The Gaussian line is consistent with an unresolved neutral Fe K$\alpha$ line in the low $\ledd$ sub-sample. The best-fitting line width in the medium $\ledd$ sub-sample is 58\,eV with an upper limit of 125\,eV. The line is not significantly broad ($<68\%$, $\Delta\chi^2=0.7$ if fixing line width at 0\,eV) and it is still consistent with an unresolved line. A broad line is only significantly detected ($>99.9\%$, $\Delta\chi^2=14.52$ if the line width is fixed at 0\,eV) in the stacked spectrum of the high $\ledd$ sub-sample. Below we fit the broad line profile with relativistically broadened line models as well as other models.

\begin{table}
  \begin{center}
    \caption{\label{tab:higedd_fit}Spectral fit of the high $\ledd$ sub-sample}
\begin{tabular}{lccc}\hline
 Parameter              & \multicolumn{1}{c}{\texttt{relline}} & \texttt{reflionx} & \texttt{double gaussian} \\\hline
$\Gamma$                & $2.06\pm0.05$                          & $2.10^{+0.11}_{-0.07}$  & $2.03\pm0.05$    \\[0.5mm]
$E_\mathrm{NL}$\,(keV)              & $6.46\pm0.10$                          & $6.48^f$                & $6.41\pm0.08$    \\[0.5mm]
EW$_\mathrm{NL}$\,(eV)& \multicolumn{1}{c}{$59^{+44}_{-45}$} & $46^{+62}_{-44}$        & $81^{+38}_{-37}$ \\[0.5mm]
$E_\mathrm{BL}$\,(keV)              & $6.90^{+0.14}_{-0.16}$                 &    ---                  &    ---           \\[0.5mm]
spin ($a$)              & \multicolumn{1}{c}{$<0.63$}            & $<0.35$                 &    ---           \\[0.5mm]
$\xi$                   &     ---                                & $49^{+1125}_{-33}$      &    ---           \\[0.5mm]
EW$_\mathrm{BL}$\,(eV)                  & $372^{+284}_{-230}$                    &    ---                  &    ---           \\[0.5mm]
EW$_\mathrm{6.67\,keV}$\,(eV) &     ---                                &    ---                  & $47^{+38}_{-45}$ \\[0.5mm]
EW$_\mathrm{6.97\,keV}$\,(eV) &     ---                                &    ---                  & $112^{+115}_{-110}$ \\[0.5mm]
$\sigma_\mathrm{6.97\,keV}$\,(eV)          &     ---                                &    ---                  & $206^p$             \\[0.5mm]
$\chi^2/\mathrm{d.o.f}$ & 77.99/73                               & 77.63/74                & 80.01/73    \\\hline
\end{tabular}
\parbox{80mm}{$f$: Parameter is fixed at its best-fitting value.\\
              $p$: Parameter pegged  at hard limit.}
\end{center}
\end{table}

\begin{figure}
  \begin{center}
    \includegraphics[width=\columnwidth]{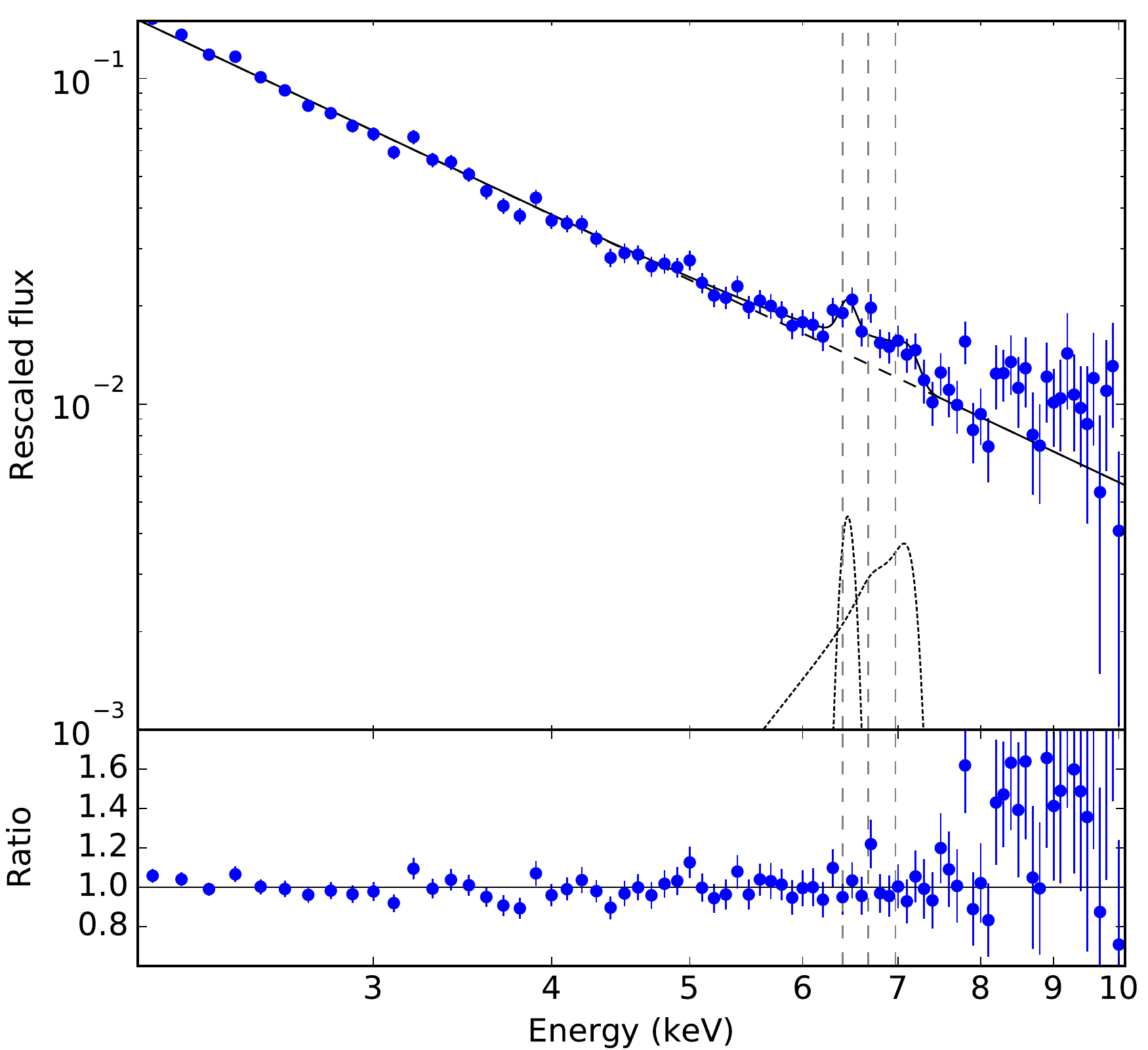}
    \caption{\label{fig:higedd}Spectral fit to the stacked spectrum of the high $\ledd$ with the \texttt{gsmooth$\otimes$(po+gaussian+relline)} model. The residuals as the data to model ratios are shown in the lower panel.}
  \end{center}
\end{figure}

\begin{figure}
  \begin{center}
    \includegraphics[width=\columnwidth]{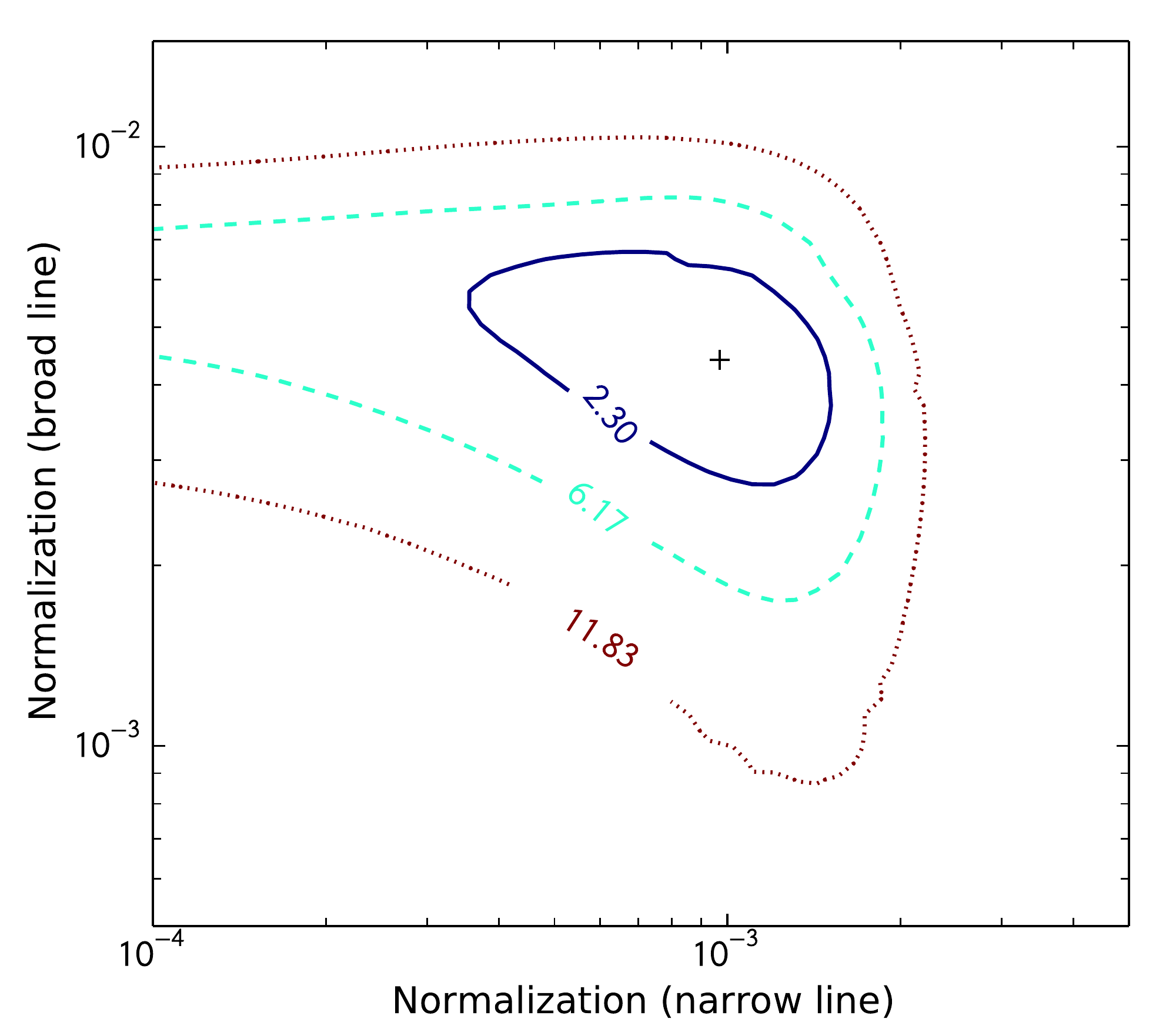}
    \caption{\label{fig:bl_nl_conf}Contours at approximate 1 (solid), 2(dashed) and 3 (dotted) sigma of the normalization of the narrow and relativistic components in the \texttt{gsmooth$\otimes$(po+gaussian+relline)} model for the high $\ledd$ sample.}
  \end{center}
\end{figure}

The large line width of the Gaussian component, i.e. $>219$\,eV, in the stacked spectrum of the high $\ledd$ sub-sample is an indication of a relativistic broad \FeK line. We fit the stacked spectrum with a broad line model by adding a \texttt{relline} component \citep{dauser:2010} to the baseline model, i.e. \texttt{gsmooth$\otimes$(po+gaussian+relline)} in \textsc{xspec} (referred to as Case A hereafter). The line width of the narrow line in the baseline model is fixed at 1\,eV. The line energy, spin and the normalization parameters in the \texttt{relline} model are free parameters. All the other parameters are fixed at the default values. The inner radius of the accretion disc is fixed at the ISCO for the current value of BH spin $a$ (i.e. the value of the \texttt{Incl} parameter in the \texttt{relline} model equals to $-1$, assuming the disc extends down to the ISCO), while the outer radius of the accretion disk is fixed at $400\,r_\mathrm{g}$. We assumed that the disc is illuminated by an isotropic primary source, thus the value of the \texttt{limb} parameter is fixed at 0. The emissivity index of the disc is fixed at 3. The inclination of the accretion disc cannot be well constrained. As most type 1 AGNs are believed to have an inclination angle in the range of $10-60\degr$, and thus we fix the inclination angle at $30\degr$ in the fitting. This model can fit the data well ($\chi^2/\mathrm{d.o.f}=77.99/73$, see Table \ref{tab:higedd_fit} and Fig. \ref{fig:higedd}). The broad line component is significantly detected ($>99.9$, $\Delta\chi^2=18.14$ if the flux of the \texttt{relline} component if fixed at 0). In Fig. \ref{fig:bl_nl_conf}, the $\Delta\chi^2$ contours for the normalization parameters of the \texttt{gaussian} and \texttt{relline} components are shown. The narrow and broad line components can only be simultaneously constrained at $1\sigma$ confidence level. The zero value of the \texttt{relline} component is excluded at $3\sigma$ level, however. The low significance may be due to the weak EW of the narrow line component ($59^{+44}_{-45}\,\mathrm{eV}$, see below). The best-fitting line energies of the narrow and broad line components are $E_\mathrm{NL}=6.46\pm0.10\,\mathrm{keV}$ and $E_\mathrm{BL}=6.90^{+0.14}_{-0.16}\,\mathrm{keV}$, respectively. The spin parameter cannot be well constrained and only an upper limit, $a<0.63$, can be given. The flux of the \texttt{relline} is $4.4^{+2.0}_{-1.8}(^{+1.6}_{-1.1})\times10^{-3}\,\mathrm{photons\,cm^{-2}\,s^{-1}}$ (the $1\sigma$ uncertainty is quoted in the parenthesis). The equivalent widths of the narrow and broad lines are $\mathrm{EW_{NL}}=59^{+44}_{-45}\,\mathrm{eV}$ and $\mathrm{EW_{BL}}=372^{+284}_{-230}\,\mathrm{eV}$, respectively. The equivalent width of the narrow line component is slightly weaker than that in the low and medium $\ledd$. This is consistent with the anti-correlation between the EW of the narrow \FeK line and the $\ledd$ found in previous stuedies \citep[e.g.][]{bianchi:2007, shu:2010, ricci:2013a}. The best-fitting line energy of the broad line may be affected by the inclination of the accretion disc. We thus fit the spectrum with the disc inclination fixed at a range of values, from $15\degr$ to $60\degr$ with a step of $5\degr$. We find that the best-fitting line energy is consistent with a neutral \FeK line only if the disc inclination is larger than $45\degr$. However, it is unlikely that the average disc inclination for type-1 AGN will be larger than $45\degr$ \citep[e.g.][]{marin:2014}, assuming that the axes of the accretion disc and the torus are aligned. Thus a highly ionized accretion disc is preferred in this work. Theoretical calculations have shown that the accretion disc may be significantly ionized for large values of accretion rate \citep{matt:1993, ross:1993, nayakshin:2000}. Observation evidences for an ionized accretion disc have also been reported in previous studies, especially for AGN with high $\ledd$ \citep[e.g. NLS1s,][]{ballyantyne:2001, liu:2015}, though most of the sources with broad line detection are consistent with a neutral disc. Adding one more narrow Gaussian component to the model (to account for the emission line feature peaking at $\sim6.7\,\mathrm{keV}$, see Fig. \ref{fig:higedd}) does not improve the fitting significantly ($\Delta\chi^2=1.57$ with 2 d.o.f).

As a more self-consistent approach, we fit the data with a smeared ionized reflection model by convolving the \texttt{reflionx} \citep{ross:2005} model with \texttt{relconv} \citep{dauser:2010}, i.e. \texttt{gsmooth$\otimes$(po+gaussian+relconv$\otimes$reflionx}). The line width of the narrow line is fixed at $1\,\mathrm{eV}$. The line energy can not be well constrained and thus is fixed at its best-fitting value $E_\mathrm{NL}=6.48\,\mathrm{keV}$. The iron abundance parameter in the \texttt{reflionx} is fixed at solar value, while the photon index of the illuminating spectrum is linked to the value of the power-law continuum. Only the BH spin is free parameter in the \texttt{relconv} component. This model can fit the data well ($\chi^2/\mathrm{d.o.f}=77.63/74$, see Table \ref{tab:higedd_fit}). The best-fitting ionization parameter is $\xi=49^{+1125}_{-33}\,\mathrm{erg\,cm\,s^{-1}}$. The equivalent width of the neutral narrow line is $\mathrm{EW_{NL}}=46^{+62}_{-44}\,\mathrm{eV}$. A low BH spin is still required, i.e. $a<0.35$.

The average spectrum can also be fitted by adding two Gaussian components to the base line model (see Table \ref{tab:higedd_fit}), i.e. \texttt{gsmooth$\otimes$(po+gaussian+gaussian+gaussian)} (referred to as Case B hereafter). The central line energies of the two Gaussian components are fixed at 6.67 and 6.97\,keV, respectively. The line widths of the neutral and 6.67\,keV Fe K$\alpha$ line are fixed at 1\,eV. The best-fitting line width of the 6.97\,keV line is 206\,eV.

\subsubsection{\label{subsubsec:nlbl_subs}The NLS1 and BLS1 sub-samples}

The base line model cannot fit the spectrum of the BLS1 sub-sample well. This is mainly due to features at low energy band ($E\lessapprox4\,\mathrm{keV}$) which can be introduced by the stacking method, as pointed out in \citet{corral:2008}. The best-fitting line width is $57_{ -57}^{ +61}\,\mathrm{eV}$, consistent with an unresolved line. No significant broad emission feature is found in the $6-7\,\mathrm{keV}$ energy band. The equivalent width of the narrow emission line is  $100_{-25}^{+27}\,\mathrm{eV}$, in agreement with previous studies (e.g. \citealt{corral:2008, zhou:2010, shu:2010, falocco:2012,  falocco:2013}).

A highly ionized emission line peaking at around $6.9\,\mathrm{keV}$ is clearly shown in the stacked spectrum of the NLS1 sub-sample (left panel Fig. \ref{fig:nl_bl}). Thus a second Gaussian component is added to the base line model, i.e. \texttt{gsmooth$\otimes$(po+gaussian+gaussian)}. The line width of the neutral line is fixed at 1\,eV. This model can fit the data well ($\chi^2/\mathrm{d.o.f}=73.25/73$). The best-fitting central energies of the neutral and ionized Fe lines are $6.44\pm0.09$ and $6.93^{+0.10}_{-0.42}\,\mathrm{keV}$, respectively. The best-fitting line width of the highly ionized emission line is 156\,eV with a lower limit of 22\,eV. The equivalent widths of the neutral and highly ionized emission lines are $103^{+62}_{-56}$ and $207^{+117}_{-107}\,\mathrm{eV}$, respectively.

\begin{figure}
  \begin{center}
    \includegraphics[width=\columnwidth]{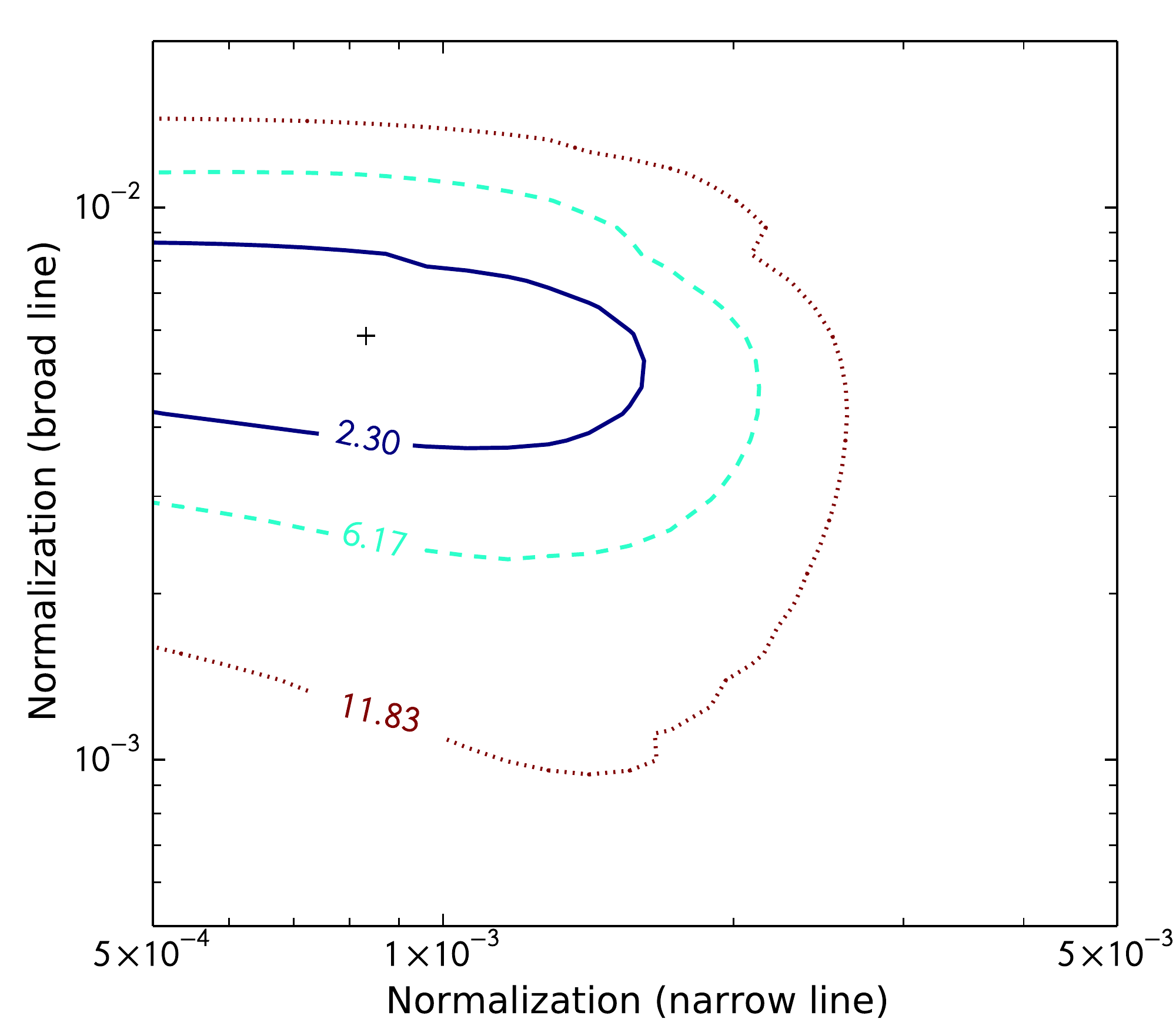}
    \caption{\label{fig:nls1_conf}Contours at approximate 1 (solid), 2(dashed) and 3 (dotted) sigma of the normalization of the narrow and broad lines for the NLS1.}
  \end{center}
\end{figure}

The line width of the highly ionized emission line in the \texttt{gsmooth$\otimes$(po+gaussian+gaussian)} model is $\sigma>22\,\mathrm{eV}$. This may be an indication of a broad \FeK line (not significant, $\Delta\chi^2=2.9$ for an additional degree of freedom). \citet{liu:2015} found a prominent broad \FeK line in the average X-ray spectrum of a NLS1 sample selected from the catalogues built by \citet{zhou:2006} and \citet{veron-cetty:2006}. We try to fit the spectrum with a relativistic broad line model, i.e. \texttt{gsmooth$\otimes$(po+gaussian+relline)}. The line energy of the narrow line cannot be constrained, thus it is fixed at its best-fitting value $E_\mathrm{NL}=6.43\,\mathrm{keV}$. The line energy and the spin are free parameters in the \texttt{relline} model. This model can fit the data well ($\chi^2/\mathrm{d.o.f}=71.79/73$). The $\Delta\chi^2$ contours for the normalization parameter of the narrow and broad line components are shown in Fig. \ref{fig:nls1_conf}. It is clear that the two components cannot be well constrained simultaneously. This may be due to the much lower S/N of the NLS1 sub-sample (see Table \ref{tab:sub_samples}) and the low EW of the narrow line component ($48^{+56}_{-48}\,\mathrm{eV}$). The spin parameter is found to be $a<0.53$. The best-fitting line energy of the broad line is $E_\mathrm{BL}=6.86^{+0.16}_{-0.12}\,\mathrm{keV}$, which may correspond to the highly ionized \ion{Fe}{xxvi} emission line. The EWs of the narrow and broad lines are $48^{+56}_{-48}$ and $495^{+288}_{-257}\,\mathrm{eV}$, respectively. These results are in agreement with the results found in \citet{liu:2015}.


\section{Discussion}

\subsection{Dependence of the broad \FeK line on $\ledd$?}\label{subsec:edd_fe}

A broad line feature, which is detected at $>3\sigma$ level ($\Delta\chi^2=18.14$ with one additional d.o.f, see Section \ref{subsubsec:edd_subs}) and can be well fitted with a relativistic \FeK line model (Case A, see Section \ref{subsubsec:edd_subs}), is revealed in the high $\ledd$ sub-sample, while no significant broad line features are found in the low and medium $\ledd$ sub-samples. The broad line feature in the high $\ledd$ sub-sample can also be fitted with multiple Gaussian components (Case B, see Section \ref{subsubsec:edd_subs}). The line energy of the broad line in Case A is consistent with a highly ionized \FeK line, assuming that the average inclination of the accretion disc is less than $45\degr$ for type-1 AGN (see Section \ref{subsubsec:edd_subs}). There is an indication that the line width of the \FeK line becomes broader as the $\ledd$ increases, as shown in Fig. \ref{fig:edd_subs} and Table \ref{tab:fit_subs}, consistent with the results presented in \citet{inoue:2007}. We should note that the relation between the $\ledd$ and the properties of the broad \FeK line may be even stronger considering that the uncertainties in the estimates of $\ledd$ can potentially weaken the relation.

A highly ionized emission line is shown in the average spectrum of a sample with high $\ledd$ in \citet[][see also \citealt{liu:2015}]{iwasawa:2012a}. As suggested by \citet{iwasawa:2012a}, the narrow ionized emission line found in their sample can be produced by outflow/winds from accretion disc. This can also explain the highly ionized emission line shown in Case B of our high $\ledd$ sub-sample. However, the highly ionized line ($E_\mathrm{BL}\approx6.94\,\mathrm{keV}$, corresponding to \ion{Fe}{xxvi}) should be originated from the accretion disc if it is a relativistic broad line as in Case A. In this case the results can be interpreted by the fact that the ionization state of the disc will be higher when the $\ledd$ goes up \citep{inoue:2007}. Theoretical calculations have shown that the ionization state of the accretion disc becomes higher as the mass accretion rate increases, producing highly ionized Fe K emission line, e.g. \ion{Fe}{xxv/xxvi} \citep{matt:1993, ross:1993, nayakshin:2000}. The equivalent width of the fluorescence line will change with the ionization state of the accretion disc: due to the resonant trapping opacity, the equivalent width of the Fe K line will first decrease, then increase strongly when the iron is more ionized than \ion{Fe}{xxiii}, reaching a maximum value of $\sim500$\,eV for very high ionization state (e.g. $\xi\sim2000$). This is consistent with the measured equivalent width of the broad line in the stacked spectrum of the high $\ledd$ sub-sample. On the contrary, the equivalent width of the Fe K$\alpha$ line will be small, e.g. $<200\,$eV for a neutral or low-ionization accretion disc, which can explain the non-detection of the broad line in the low and medium $\ledd$ sub-samples (see Sec. \ref{subsec:nl_bl}). 

Alternatively, in case A the results can also be interpreted if the truncation radius of the accretion disc is strongly dependent on the $\ledd$. It has been suggested that the optically thick physically thin accretion disc, which exists at the low $\ledd$ regime, terminates beyond the ISCO and the inner region is filled with a hot advection-dominated accretion flow\citep[ADAF, e.g.][]{narayan:1994, narayan:1995}. This model has been widely used as a mechanism for explaining state changes in X-ray binaries \citep[XRB, e.g.][]{esin:1997, done:2007}. Spectral evidence for disc truncation has found in XRB \citep[e.g.][]{esin:2001, done:2010} and some radio-loud AGN \citep{marscher:2002, lohfink:2013}. The non-detection of the relativistic line is expected if indeed the disc is truncated at larger radius for AGN with low $\ledd$, as the case for our low and medium $\ledd$ sub-samples. On one hand, the profile of the emission line, as well as the reflection spectrum will not be significantly blurred by the relativistic effect at large radius. On the other hand, both the emissivity index, which is thought to be larger at the inner part of the accretion disc, and the region that is illuminated by the power-law continuum will decrease if the accretion disc is terminated at large radius, thus the strength and the equivalent width of the emission line will be weak.

\subsection{Broad \FeK line in NLS1 and BLS1 galaxies}\label{subsec:nl_bl}

\begin{figure*}
\begin{center}
    \includegraphics[width=1.0\columnwidth]{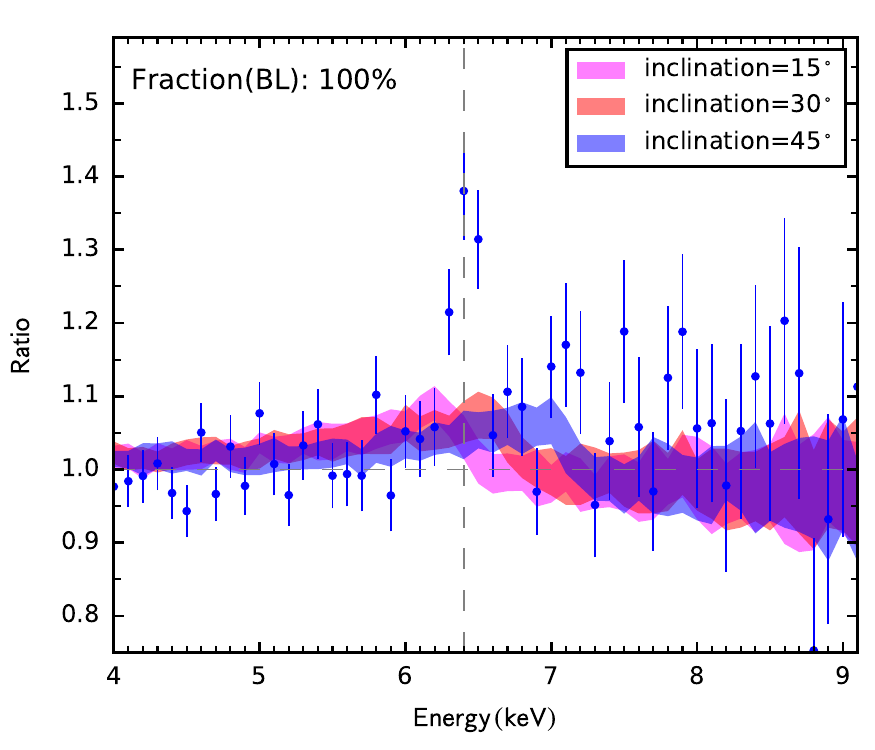}
    \includegraphics[width=1.0\columnwidth]{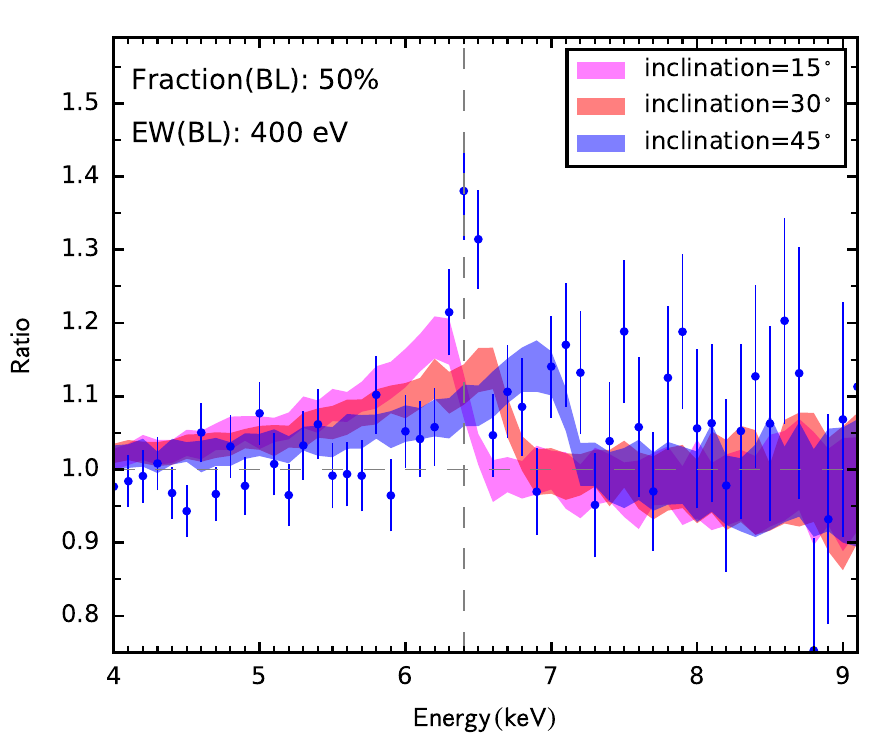}
    \caption{The ratio of the stacked spectrum to a best-fitting continuum for the BLS1 sub-sample is shown as blue points. Different colours represent different disc inclinations (magenta: $i=15\degr$; red: $i=30\degr$; blue: $i=45\degr$). Left: the shadow areas show the $1\sigma$ confidence interval of the stacked spectra for the simulated sources (fraction $f=100$). The distribution of the EW for the broad line is estimated based on the $\sim44$ sources which have broad \FeK line measurements. Right: the EW of the broad \FeK line in the simulations is fixed at 400\,eV, while the fraction of sources with broad \FeK line is 50 per cent. \label{fig:bl_percent}}
\end{center}
\end{figure*}
In addition to the neutral \FeK line, a highly ionized emission line feature is clearly shown in the average spectrum of the NLS1 sub-sample (Fig. \ref{fig:nl_bl}). The best-fitting line width of the highly ionized emission line is 156\,eV with a lower limit of 22\,eV. This may be an indication of a broad \FeK line, though with low significance ($\Delta\chi^2=2.9$ if the line width is fixed at 0). The broad relativistic \FeK line should be common in NLS1, as suggested by \citet{liu:2015}. They found a prominent broad \FeK line in the composite X-ray spectrum of a NLS1 sample consisting of 51 sources. As shown in the left panel of Fig. \ref{fig:mbh_edd}, there is a large overlap between the NLS1 and high $\ledd$ sub-samples, thus it is not surprising that a broad line may be detected in the NLS1s if the properties of the broad line indeed depends on $\ledd$ as suggested in Section \ref{subsec:edd_fe}. The stacked spectrum of our NLS1 sub-sample can be well fitted with the relativistic line model. The low significance of the broad line in our NLS1 sub-sample may be due to the much lower S/N of the data (see Table \ref{tab:sub_samples}). The EWs of the narrow ($48^{+56}_{-48}\,\mathrm{eV}$) and broad line ($495^{+288}_{-257}\,\mathrm{eV}$), as well as the BH spin ($a<0.53$), are in agreement with the results found in \citet{liu:2015}.

No broad line feature is detected in the average spectrum of our BLS1 sub-sample and also the whole sample. \citet[][see also \citealt{corral:2008, falocco:2012, falocco:2013}]{falocco:2014} found a significant relativistic line in the stacked spectrum of a sample selected from the AGN catalogue built by \citet{veron-cetty:2010}, however. This may be simply due to the different samples used in different studies. The samples used in those previous works consist of different types of AGN, and often have very different properties (e.g. higher redshift and different luminosity range) comparing with our sample. Here we compared the results between the our BLS1 and NLS1 sub-samples, as well as the NLS1 sample presented in \citet{liu:2015}.

The non-detection of the broad line in the BLS1 sub-sample can be explained if the fraction of sources with relativistic \FeK line is lower, or the average EW of the broad line is relatively smaller, than that in the NLS1 ($\sim 400$\,eV found in \citealt{liu:2015}). To roughly estimate the fraction of sources with broad \FeK line in BLS1 galaxies we carry out simulations. Using the method described in Sec.\,\ref{subsec:simulations}, we simulate a set of `observed' spectra for the sources in the BLS1 sub-sample. We assume that the fraction of sources with broad line is $f$ in the simulated spectra of the BLS1 sub-sample. Spectra without a relativistic component are generated using the best-fitting continua of the sources, while for the remaining ones a broad line model is added to the best-fitting continua. The spin parameter is randomly distributed in the range of 0-0.998. In order to compare the average broad line profile for different inclination angles, we simulate spectra for three disc inclinations, i.e. $i=15\degr$, $30\degr$, $45\degr$. Two different distributions for the EW of the relativistic component are assumed. In the first case (referred as Case 1 hereafter) the line EWs are randomly drawn from a probability distribution constructed on the basis of the $\sim 44$ sources \citep[][two sources with very strong broad \FeK, i.e. EW$>600$\,eV, are excluded]{brenneman:2013} which have broad \FeK line measurements. The typical values for the EW of the broad line component in Case 1 are smaller than the average line EW found in the stacked spectrum of our NLS1 sub-sample, e.g. $<300$\,eV. In another case (Case 2) the EW of the broad line component in the simulation is fixed at a comparable value to the observed average EW of the broad \FeK line in NLS1, e.g. $\mathrm{EW}=400$\,eV.

The same stacking method is then applied to those simulated spectra. The results for Case 1 and Case 2 are shown in the left and right panel of Fig. \ref{fig:bl_percent}, respectively. The shadow areas represent the $1\sigma$ confidence intervals for the simulated composite spectra. The observed data of our BLS1 sub-sample are shown with blue points. The different colours represent the stacked line profiles for different disc inclinations (magenta: $i=15\degr$; red: $i=30\degr$; blue: $i=45\degr$). It is obvious that the non-detection of the broad \FeK line is not surprising in Case 1, since the observed data is consistent with the simulated results for all the different disc inclinations even when $f$ equals to 100 per cent. The fraction of sources with broad \FeK line in our BLS1 sub-sample should be much smaller than 50 per cent in Case 2 if the disc inclination is $i=15\degr$. For larger disc inclinations, e.g. $i=30\degr$ (red) and $45\degr$ (blue), the simulated results are consistent with the observed data if fraction $f\la50$\,per cent. This fraction is compatible with the fractions reported in previous studies, e.g.  $\sim45\,$per cent in \citet{nandra:2007}; $36\,$per cent in \citet{de-la-calle-perez:2010} and $\sim50\,$per cent in \citet{patrick:2012}.

\subsection{Non-detection of broad line in the $\mobh$ sub-samples}

No significantly broad line feature is detected in the stacked spectra of the $\mobh$ sub-samples, though an indication of broad line in the MBH3 sub-sample is present. This may suggest that the fraction of sources with broad \FeK line or/and the equivalent width of the broad \FeK line do not strongly depend on the black hole mass, considering that a broad line is detected in the high $\ledd$ sub-sample with similar S/N. The non-detection of broad \FeK line in the $\mobh$ sub-samples can also be understood if the EW of the broad line is mainly driven by the $\ledd$. As shown in Fig. \ref{fig:mbh_subs}, although the $\mobh$ is slightly correlated with $\ledd$, there is a large dispersion of the $\ledd$ distribution in each $\mobh$ bin, which leads to a relatively small average EW of the broad line, thus make it difficult to be detected.

\section{Summary}

In this paper, we use a large sample of AGN, which have well measured optical parameters, to investigate the dependence of the broad \FeK line on the physical parameters of AGN, such as the $\mobh$, $\ledd$, and optical classification, by means of X-ray spectral stacking. A broad \FeK line feature is detected ($>3\sigma$) in the high $\ledd$ sub-sample ($\log\ledd>-0.9$). Our results indicate a dependence of properties of the broad \FeK line on the $\ledd$. Indications of the broad line is also found in the sub-samples with $\log\mobh\sim7.84\,M_{\sun}$ and the NLS1 sub-sample, though with low significance.  No significant broad line feature is shown in the stacked spectra of the BLS1 sub-sample. This may suggest that the fraction of sources with broad \FeK line is low or the EW of the broad \FeK line is small in BLS1 galaxies, comparing with the NLS1 galaxies. Future large X-ray samples, such as those selected with eROSITA/XTP, are needed to improve the results.

\section*{Acknowledgements}
This work is supported by the National Natural Science Foundation of China (grant No.11473035, 11273027, 11303046), and the Strategic Priority Research Program “The Emergence of Cosmological Structures” of the Chinese Academy of Sciences (grant No. XDB09000000). XBD acknowledges financial support through NSF grant 11473062. FJC acknowledges financial support through grant AYA2015-64346-C2-1-P (MINECO/FEDER)
This work is based on observations obtained with XMM-Newton, an ESA science mission with instruments and contributions directly funded by ESA Member States and NASA. 










\bsp	
\label{lastpage}
\end{document}